\documentclass[10pt]{iopart}
\usepackage{epsfig}
\newcommand{\be}{\begin{eqnarray}}
\newcommand{\ee}{\end{eqnarray}}

\newcommand{\ave}[1]{\left\langle #1 \right\rangle}

\newcommand{\order}[1]{ \mathcal{O} \left( #1 \right) }

\begin{document}
\title{Statistical hadronization phenomenology in $K/\pi$ fluctuations at ultra-relativistic energies}
\author{Giorgio Torrieri$^{a}$,Rene Bellwied$^b$,Christina Markert$^c$,Gary Westfall$^{a,d}$}
\address{$^a$ FIAS,
  J.W. Goethe Universit\"at, Frankfurt A.M., Germany.}
\address{$^b$ Wayne State University, Detroit, Michigan 48201, USA}
\address{$^c$ University of Texas at Austin, Austin, Texas 78712, USA}
\address{$^d$ Michigan State University, East Lansing, Michigan 48824, USA}
\maketitle
\begin{abstract}
We discuss the information that can be obtained from an analysis of fluctuations in heavy ion collisions within the context of the statistical model of particle production.  We then examine the recently published experimental data on ratio fluctuations, and use it to obtain constraints on the statistical properties (physically relevant ensemble, degree of chemical equilibration, scaling across energies and system sizes) and freeze-out dynamics (amount of reinteraction between chemical and thermal freeze-out)  of the system.  
\end{abstract}
The idea of modelling the abundance of hadrons using statistical mechanics techniques has a long and distinguished history ~\cite{Fer50,Pom51,Lan53,Hag65}.  In a sense, any discussion of the thermodynamic properties of hadronic matter (e.g. the existence of a phase transition) {\em requires} that statistical mechanics be applicable to this system ( through not necessarily at the freeze-out stage).

That such a model can describe {\em quantitatively} the yield of most particles, including multi-strange ones, has in fact been indicated by fits to average particle abundances at AGS,SPS and RHIC energies  ~\cite{bdm,equil_energy,jansbook,becattini,nuxu,share}.

There are, however, several points of contention to this understanding:  Some practitioners,starting from \cite{Fer50} interpret the statistical model results in terms of nothing more than phase space dominance:  For a process strongly enough interacting with enough particles in the final state, dynamics ``factors out'' into a normalization constant, and the final state probabilities are dominated by phase space.  If this is the case, the applicability of the statistical model has nothing to do with a genuine equilibration of the system.

Others \cite{bdm,jaki} believe that the applicability of the statistical model is a sign of a phase transition, as the chemical equilibration of hadrons signals a regime in which multi-particle processes and high-lying resonances dominate.

Still others think that in soft QCD processes particles are ``born in equilibrium'' \cite{castorina}, and the applicability of the statistical model to even smaller systems (itself contested \cite{pbmel}) is a fundamental characteristic of QCD.

How can these scenarios be differentiated?  One observable ``at the heart'' of statistical mechanics is  event-by-event fluctuations \cite{prcfluct,sharev2}.
It is a fundamental principle of statistics that variances around averages scale a certain way w.r.t. averages.  In our context ``Averages'' are particle multiplicities per event and fluctuations are event-by-event fluctuations.   For macroscopic systems, this principle ensures that fluctuations become negligible and the expectation that the state of the system is the maximum entropy one is nearly certain to be realized.

$\order{100-1000}$ particles is not enough for this to be the case, but, if statistical mechanics applies, one should still see that yields, fluctuations and higher cumulants scale in a way calculable from the partition function.   There are,however, some experimental issues specific to heavy ion collisions that need to be explored before this can be transformed into a quantitative test.

``standard'' fluctuation calculations assume that the volume\footnote{The applicability and definition of the term ``volume'' is itself problematic in a dynamical quantum system.   Statistical particle production {\em assumes} a volume, but leaves unexplained how this volume is defined, eg if acceptance cuts can be considered as ``cuts in volume''.  If particles are created with a boost-invariant flow,$v_z=z/t$, and the thermal parameters are approximately independent of rapidity, than cuts in rapidity are equivalent to cuts in volume.  No other kinematic cuts, eg cuts in $p_T$ or $\phi$, can be reliably treated in this way, although the effect of these cuts on fluctuations can be {\em partially} cancelled out by mixed event subtraction, as discussed later in this work} of the system is fixed.  Obviously, in heavy ion collisions, the system is not in a box, so this requirement has no reason to hold.   The most straight-forward way around this is to construct an ensemble where volume can fluctuate while its canonical conjugate, the pressure, stays fixed \cite{hagpress}.  This, however, can also present phenomenological difficulties, since the origin of volume fluctuations in heavy ion collisions is not necessarily statistical, but also geometric (initial state fluctuations) and dynamical (non-zero Knudsen number \cite{v2fluct}).   Statistical mechanics, therefore, must include these fluctuations in the constraints, since it has no ability to describe them.

Due to our incomplete understanding of how the initial state and dynamics contribute to fluctuations at freeze-out, the best approach is to choose an observable which is {\em insensitive} to any volume fluctuations.   In the thermodynamic limit, where volume becomes a proportionality constant at the level of the partition function, a tempting observable is the scaled variance of the {\em ratio} of two particle multiplicities measured event by event.
That this is in fact a good guess can be seen at the level of
particle distributions.  Assuming we are in the thermodynamic limit, the distribution should separate into a component sensitive to volume $V$ and the other on density, $N_1/V$
\begin{equation} F_{1,2}(N_{1,2}) = \int g(V) f_{1,2} \left( \frac{N_{1,2}}{V} \right) 
dV
\end{equation}
here $f_{1,2}$ depend on thermodynamics and $g(V)$ on initial geometry and dynamics (note that we left it as a general function, and in general we do not really know what this function is).
Now, the probability distribution function of a ratio is
\begin{equation}
F\left(R=\frac{N_1}{N_2}  \right)= \int F_1 (N_1) F_2 (N_2) \delta \left( \frac{N_1}{N_2} - R \right) 
\end{equation}
Expanding and substituting $\alpha = V N_2$ we get
\begin{equation}F(R) = \underbrace{\int g(V)^2 V dV}_{Independent\phantom{A}of\phantom{A}R} \int f_1 \left( \alpha R \right) f_2 \left( \alpha \right) d \alpha \end{equation}
\noindent hence, $\mathcal{N}=\int g(V)^2 V dV$ appears \textit{equally} in \textit{all} cumulants.

Thus, quantities such as $\sigma_{N_1/N_2}$ are \textit{strictly} independent of any volume fluctuations \footnote{\cite{konch} section IV-C makes the opposite claim based on an expansion around $\Delta V/\ave{V}$.  The flaw in that procedure is that it can not be determined weather terms such as $\Delta V/\ave{V}^2$ are to be included in the fluctuation or in higher cumulants.  The transport analysis in \cite{konch} validates our derivation, since strictly {\em no} dependence of ratio fluctuations on multiplicity fluctuations is found}.   Note that this is not true, for example, for the Kurtosis, recently proposed as signatures for the critical point \cite{stephanov}, since the Kurtosis is not the ratio of two cumulants.   The procedure above, however, makes it easy to derive the appropriate volume-independent observable encoding higher cumulants of ratios ($\mu_4/\sigma$ of the distribution of a ratio, for example, would be volume independent.  Here $\mu_4$ is the fourth cumulant and $\sigma$ the fluctuation).

\begin{figure}[h]
\begin{center}
\epsfig{width=14cm,figure=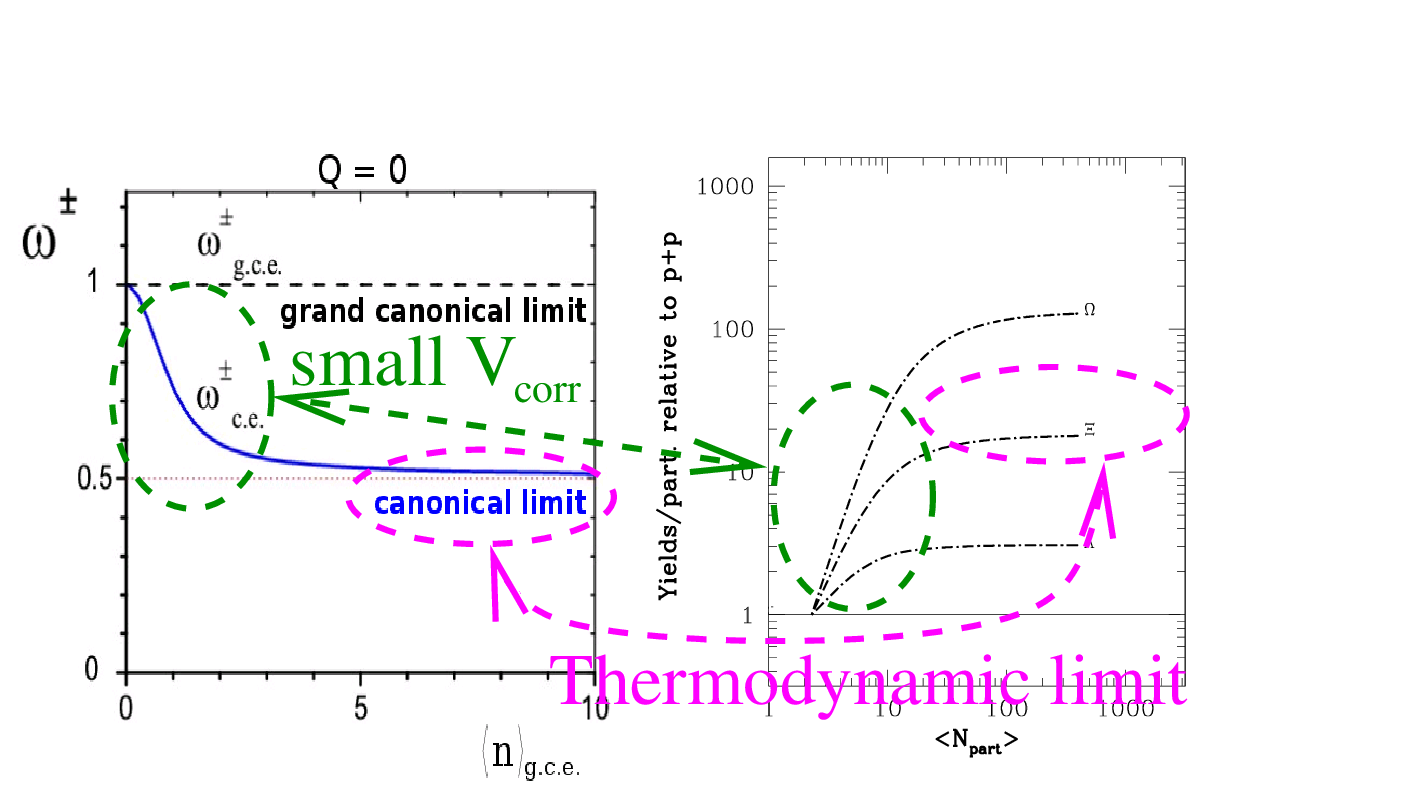}
\caption{\label{figcan} (left) Fluctuations dependence in canonical scenario \cite{canfluct} (right) Strangeness fluctuations in the canonical scenario \cite{canonical1}}
\end{center}
\end{figure}

The residual dependence of $\sigma_{N_1/N_2}$ on the average volume $\ave{V}$ can be in turn eliminated, in the grand canonical ensemble, by focusing on $\Psi=\ave{N_1} 
\sigma^{N_1/N_2}_{dyn}$, where $\ave{N_1}$ and $\sigma_{dyn}$ are to be {\em measured within the same acceptance}.  Note that this independence is specific to the grand-canonical ensemble, so should not apply to scenarios where the ``enhancement of strangeness'' in A-A collisions is due to the transition between the canonical limit in elementary processes and the grand canonical limit in A-A \cite{canonical1}.   In this scenario, the ``strangeness correlation volume'' ($V_{corr}$ not in general equal to the system volume) should regulate $\Psi$ as in Fig. \ref{figcan} \cite{canfluct}

A different problem is the effect of a detectors limited acceptance
( Particle (mis)identification, Limited rapidity and momentum resolution, momentum cuts necessary to eliminate jets etc) on fluctuation observables.  These are much more difficult to model than averages, and once again an observable needs to be constructed insensitive to them.
Hence, the necessity of mixed event subtraction \cite{methods}.   Mixed events here, are {\em defined} as events where {\em no physical correlation from the original event} are left in.  This means that any correlation seen is due to the imperfection of the detector (the fact that detector acceptance excludes particles of $p_T>1$ GeV for both event A and B creates a correlation between $A$ and $B$).
To a good approximation, examined in the next paragraph, the measured fluctuation is $\sigma^2 = \sigma_{physics}^2 + \sigma_{acceptance}^2$
and the mixed event one is $\sigma_{mix}^2 = \sigma_{trivial}^2 + \sigma_{acceptance}^2$.
Therefore, concentrating on $\sigma_{dyn}^2 = \sigma^2 - \sigma_{mix}^2$
\textit{should} eliminate acceptance effects.  

\begin{figure}[h]
\begin{center}
\epsfig{width=8cm,figure=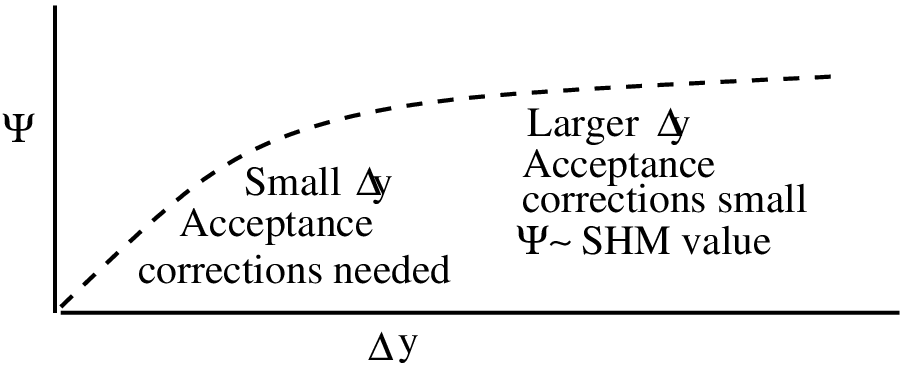}
\caption{Correlation as a function of acceptance \label{sigmadyn}}
\end{center}
\end{figure}

Two considerations are needed at this point:  The first is that this method is not perfect, since limited acceptance spoils not just fluctuations but correlations, and this will {\em not} be corrected with mixed event subtraction.
For example, in a narrow acceptance detector, resonance kinematics 
modifies $\sigma_{physics}^2$ (weakens the correlation due to the resonance) rather than introducing an additional $\sigma_{acceptance}^2$  (HBT would be another source of such physical correlations, although the smallness of the relevant relative momentum makes this largely irrelevant here).
There is no perfect solution: One empirical fix is to vary the acceptance, until its large enough that $\Psi$ reaches an asymptotic limit Fig. \ref{sigmadyn}.  This limit is the physical value of $\Psi$.  Such a method is not perfect, but it relies on the fact that, for a good event mixing procedure (defined above) limited acceptance should only {\em destroy} correlations, not {\em create} them.  The highest value of the correlation seen by varying the detector acceptance is therefore the physical one.

\begin{figure}[t]
\begin{center}
\epsfig{width=8cm,figure=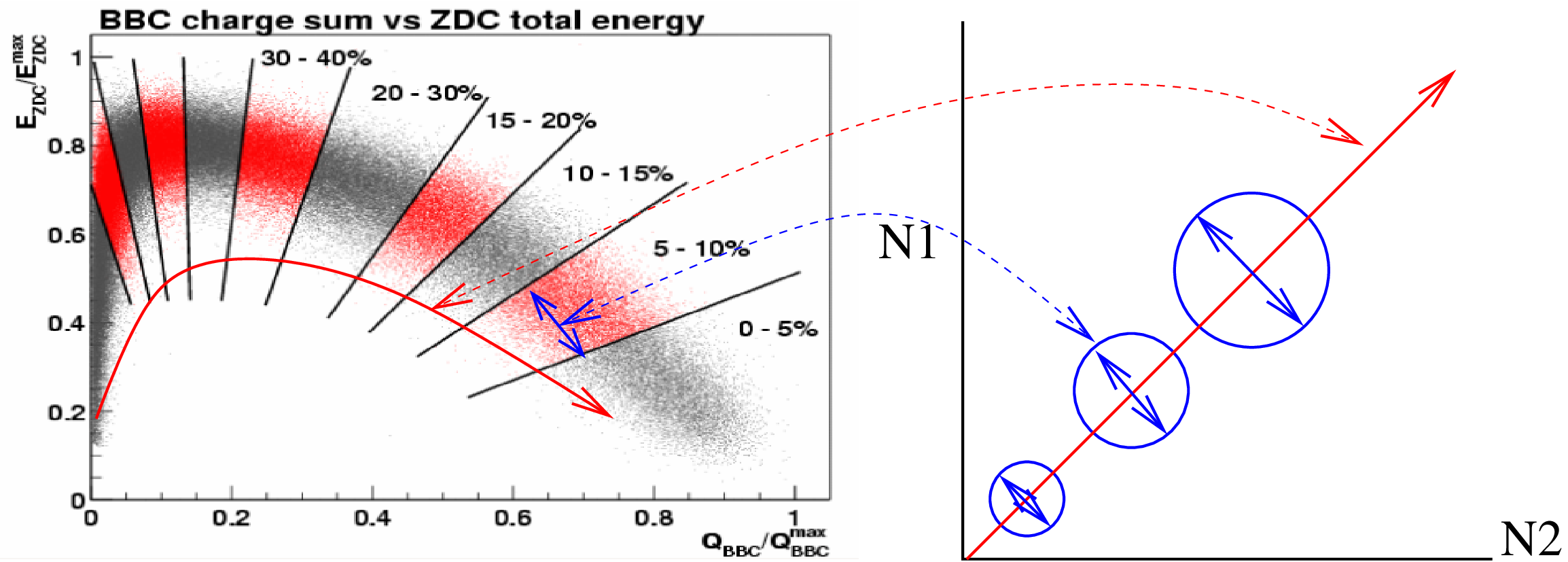}
\caption{\label{glaubermult} The experimental multiplicity measure used to normalized mixed events (left panel) and an example of the systematic correlation this procedure generates (right panel) in a ratio between two generic $N_1$ and $N_2$ particles}
\end{center}
\end{figure}

The second consideration is that, by the definition above ({\em no physical correlations} in mixed events), the current algorithms used by experimental collaborations  \cite{fluctstar,na49fluct} for mixed event definition are {\em not} perfect.  For example, the {\em total multiplicity} of mixed events is determined according to the experimentally measured multiplicity distribution, containing all correlations of real events (eg, the $\pi^+$  $\pi^-$ correlation induced by the $\rho$).    Since the abundance of all species are greater in a higher multiplicity event, the autocorrelations contained in the experimentally measured multiplicity distribution will translate into a correlation in the mixed event ratio of {\em any} two particles (see Fig \ref{glaubermult} for an illustration).  This correlation is physical rather than acceptance-induced, and hence should not be there in a ``good'' mixed event.

A more sound procedure,therefore, would be to normalize $\pi,K$ and protons {\em separately} and construct the mixed event bottom-up.     In other words, for each mixed event $\pi,K$ and p multiplicities should be determined according to the respective experimentally measured multiplicity distributions (uncorrected for acceptance).
Then the event should be constructed (the appropriate number of $\pi,K$ and proton tracks obtained from different events).
As physical autocorrelations within the the same particle species of relative momenta larger than $\order{100}$MeV  are negligible, this procedure should yield a much better (according to the above definition) mixed event sample. 
We do not think the multiplicity auto-correlation in mixed events is large, but it is there, and it could be responsible for the deviation between statistical model and the data shown in in the next paragraphs.  
It will be interesting to see how implementing the mixed event definition presented here will change the results.

\begin{figure}[h]
\begin{center}
\epsfig{width=18cm,figure=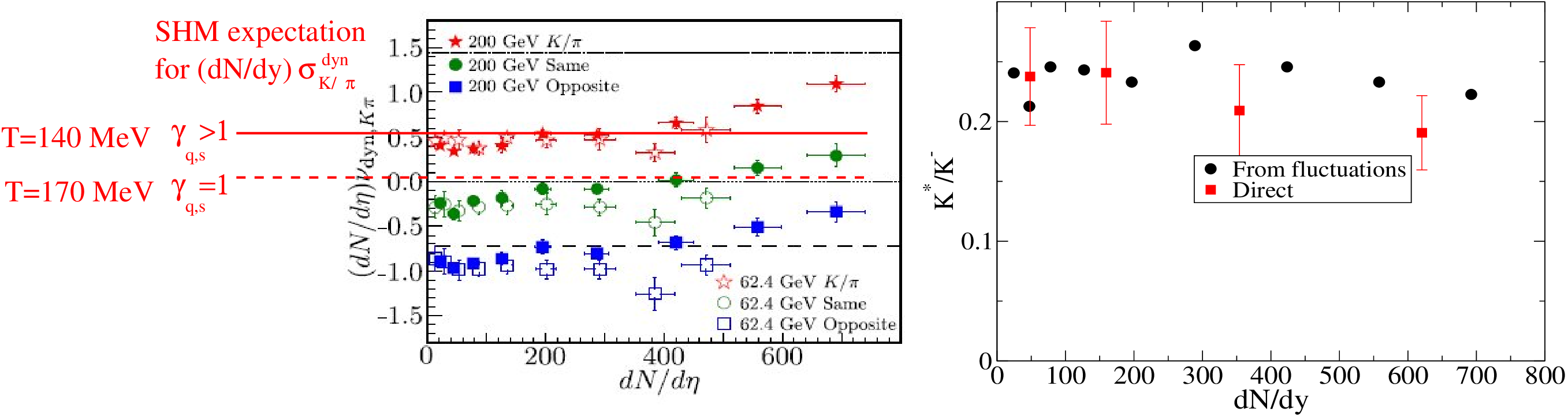}
\caption{\label{fluctstar}  Left panel:  Scaling of $K/\pi$ fluctuations at RHIC 200 GeV.  Right panel: $K^*/K$ implied from fluctuations and directly measured.
 Star fluctuations (left) and resonances (right) results, taken respectively from \cite{fluctstar,fluctreso}}
\end{center}
\end{figure}

We now proceed to discuss specific results and conclusions that can be obtained from a study of fluctuation data.
We refer to  ~\cite{bdm,equil_energy,jansbook,becattini,nuxu,share,sharev2} for a review of the statistical model.    For this work, it is sufficient to repeat that the particle abundances and fluctuations can be calculated from the first and second derivative's a particle's partition function.    In the Grand-Canonical limit, this partition function is strictly proportional to volume.   Because of this, the fluctuation $\ave{(\Delta N)^2}$  and the yield $\ave{N}$ should both scale linearly with volume, and hence the fluctuation of a ratio
\begin{equation}
\sigma_{N_1/N_2} = \frac{\ave{(\Delta N_1)^2}}{\ave{N_1}^2} + \frac{\ave{(\Delta N_2)^2}}{\ave{N_2}^2} - \frac{\ave{\Delta N_1 \Delta N_2 } }{\ave{N_1} \ave{N_2}} + \mathcal{O}\left( \frac{\Delta N}{\ave{N}^2} \right)
\end{equation}
should scale {\em inversely} to the volume $\ave{V}$ (and be strictly independent of higher cumulants of $V$, by the earlier derivation).   Hence, provided the chemical parameters do not change across systems, $\Psi_{N_1/N_2}^{N_1} = \ave{N_1} \sigma^{N_1/N_2}_{dyn}$ should be strictly independent of centrality and system size.  This requirement is not satisfied by the SPS scan \cite{na49fluct} discussed in the last part of the paper, since there $\mu/T$ does vary considerably \footnote{ A scan in centrality and system size within the same energy should however produce the same approximately horizontal bands in $\Psi$ as are seen at RHIC: Each value of $\sqrt{s}$, and hence $\mu/T$, should produce a band when scanned in centrality and system size}.   It is however satisfied by the RHIC upper energies.  Thus, the Grand-Canonical statistical model predicts a flat dependence with system size at these energies. 
  The canonical model, on the other hand, would predict a ``kink'' within the same centrality as when the strangeness correlation volume approaches the thermodynamic limit (Fig. \ref{figcan}).

The results can be seen on Fig. \ref{fluctstar} left panel.  None of the models come out perfectly, but the Grand-Canonical model is qualitatively much more similar to the data than the canonical one, as no kink is visible there.
The discrepancy in $\Psi$ between 200 GeV and 62 GeV, the slight upward trend in centrality, and the lack of scaling between A-A and Cu-Cu should however be closely watched, as the slight increase of fluctuations with multiplicity can not be accounted for by {\em any} statistical model (Since we are dealing with ratio fluctuations, a volume fluctuation tuned to KNO scaling, as in \cite{fluctvolume}, will not help here).
  
Before passing judgement on this topic, however, it will be interesting to see how much of this trend is due to the previously described multiplicity correlation retained in mixed events.   Correcting for this effect using an improved mixing procedure would go in the right direction (since physical correlations due to volume would not be cancelled out).  An analysis is ongoing to determine their extent.   The scaling of fluctuations between RHIC and the LHC, where $\mu/T$ is virtually the same, should provide further light on this.

\begin{figure}[h]
\begin{center}
\epsfig{width=13cm,figure=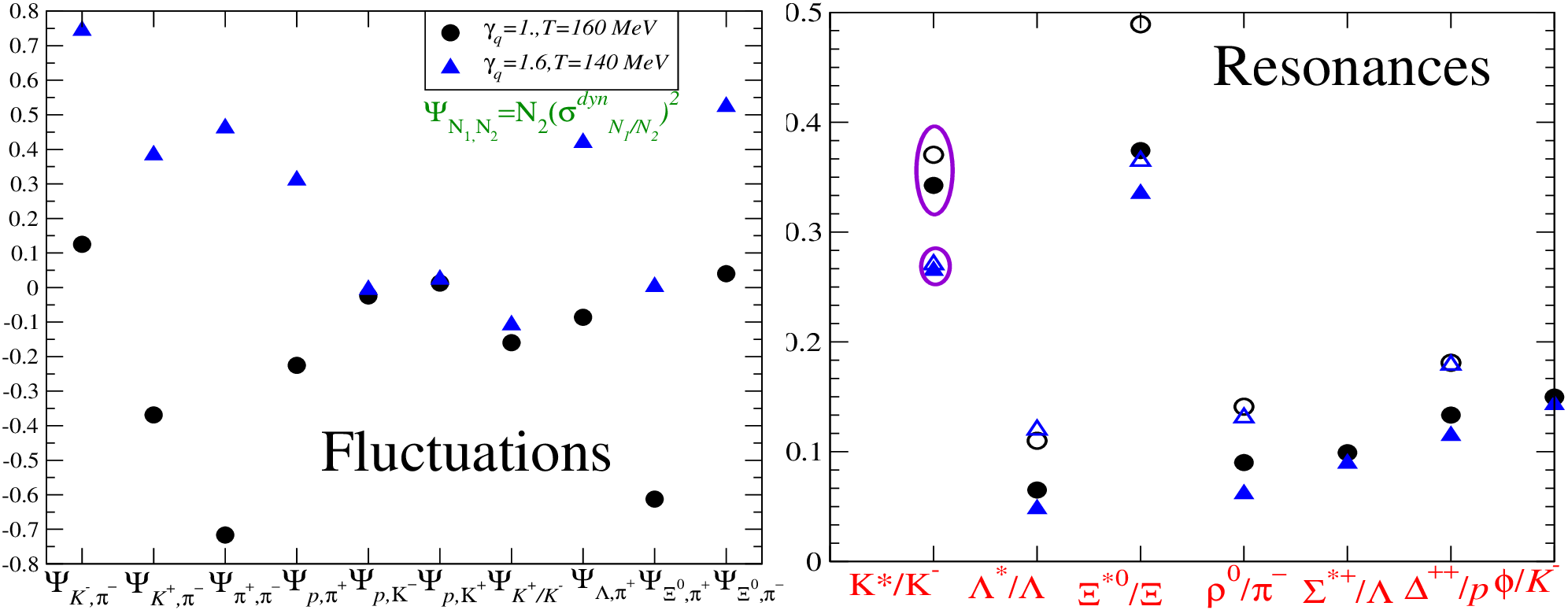}
\caption{\label{plotrhic}  Equilibrium and non-equilibrium  predictions from fluctuations (left panel) and resonances (middle panel) with the statistical parameters taken from \cite{jan_energy} (The $\gamma_q=1$ values are similar to \cite{bdm}).  In the right panel, Full symbols show the $N^*/N$ ratio inferred from the correlation in the two models, while empty symbols show, wherever possible, the estimate from $\sigma_{dyn}$ comparisons  }
\end{center}
\end{figure}

Quantitatively, as reported earlier, $\Psi$ is modelled much better with the inclusion of the light-quark non-equilibrium parameter $\gamma_q$, due to Bose-Einstein enhancement of fluctuations \cite{prcfluct}.   It remains to be seen weather the measurement of fluctuations of {\em 
more} particles (Fig. \ref{plotrhic} left panel) will corroborate this conclusion.

We now turn to a potentially important model-independent use of fluctuations, the constraint on resonance reinteraction between chemical freeze-out and thermal freeze-out \cite{jeon}.   The former can be estimated, to a good approximation, by comparing the fluctuations of same-charged particles (uncorrelated) and opposite charged particles (correlated by $K^*$) 
\begin{equation}
\frac{K^{*0}}{K^-} \simeq \frac{3}{4} \left( \Psi_{K^-/\pi^-}^{\pi^-} -  \Psi_{K^+/\pi^-}^{\pi^-}  \right)
\end{equation}
Note that both $\Psi_{K^-/\pi^-}^{\pi^-}$ and $ \Psi_{K^+/\pi^-}^{\pi^-}$ are equally affected by the volume correlations present in mixed events, so these should cancel out when the difference is taken.
Comparing the fluctuation estimate  of $\frac{K^{*0}}{K^-}$ to a direct measurement should yield the amount of $K^{0*}$ destroyed by rescattering or regenerated through pseudo-elastic interactions.

The results are shown in Fig. \ref{fluctstar} (right panel, resonance values are taken from \cite{fluctreso}).   Rescattering and regeneration have,within error bar little or no effect on the final abundance of $K^*$s.    Either chemical and thermal freeze-out proceed very close to each other, so the amount of reinteraction is negligible (as was concluded in \cite{markreso} based on \cite{ratio1,ratio2}), or rescattering and regeneration of detectable resonances cancel each other out to the degree of approximation allowed by the error bars ($10-15 \%$).    This margin appears somewhat below the estimates from transport models, which are $\sim 40\%$ \cite{urqmd1,urqmd2}. 
This discrepancy is not too surprising, since, within the framework of a collectively expanding system, a balance between rescattering and regeneration is unnatural: 
If the number of reinteractions is small, rescattering generally dominates as was assumed in \cite{ratio1,ratio2}.  If the number of rescattering is 
large, detailed balance would mean $K^*/K$ would reequilibrate from the higher chemical freeze-out to the lower thermal freeze-out temperature.

It would therefore be very interesting to investigate weather a transport model such as uRQMD \cite{urqmd3} or HSD \cite{konch} could be tuned to simultaneously reproduce the fluctuation inference and direct measurement of $\frac{K^{*0}}{K^-}$.

Unfortunately, the baroqueness of the resonance decay tree severely limits  the feasibilness of such graphic methods, as the right panel of Fig \ref{plotrhic} shows.     $K,\pi$ is a special pair of particles in that there is only one type of resonances that decays into both, the $K^*$, and its lightest state is considerably lighter than the heavier ones.   No similar definition is possible for $\Sigma^*/\Lambda,\rho/\pi$ and $\phi/K$, since the resonance decays equally into all pairs of decay products.
For other resonances, cross-contamination destroys any value of fluctuations as a {\em graphic} tool of reinteraction time.

This does not mean, however, that such resonances are useless for constraining reinteraction time, since both fluctuations \cite{prcfluct} and resonances \cite{ratio1,ratio2} provide a tight constraint on all statistical models.
If a statistical model consistently describes fluctuations, but over-predicts resonance abundances, it could be taken as an indication of a long reinteraction times with detailed balance.   A casual look at experimental data \cite{fluctreso} shows this does not seem to be the case, since most resonances are {\em under-predicted} by statistical models.

We now turn to the energy scan seen at SPS energies.  We first warn the reader not to take any quantitative statements we make here too seriously, since to study the scaling with  volume requires the same acceptance region from all particles (see footnote on page 2), as well as for the yield and fluctuation measurements.  The analysis in \cite{na49fluct} does not meet this requirement, since the acceptance for $K$ and $\pi$ is considerably different, and the acceptance for yields is different still (note that the earlier analysis in \cite{qm2007} failed to consider this).  We await for measurements where this issue is resolved (for experimental reasons they are easier to perform at collider energy scans \cite{starscan,phescan,nica})  to see how big it is of an issue for the SPS results.

\begin{figure}[h]
\begin{center}
\epsfig{width=15cm,figure=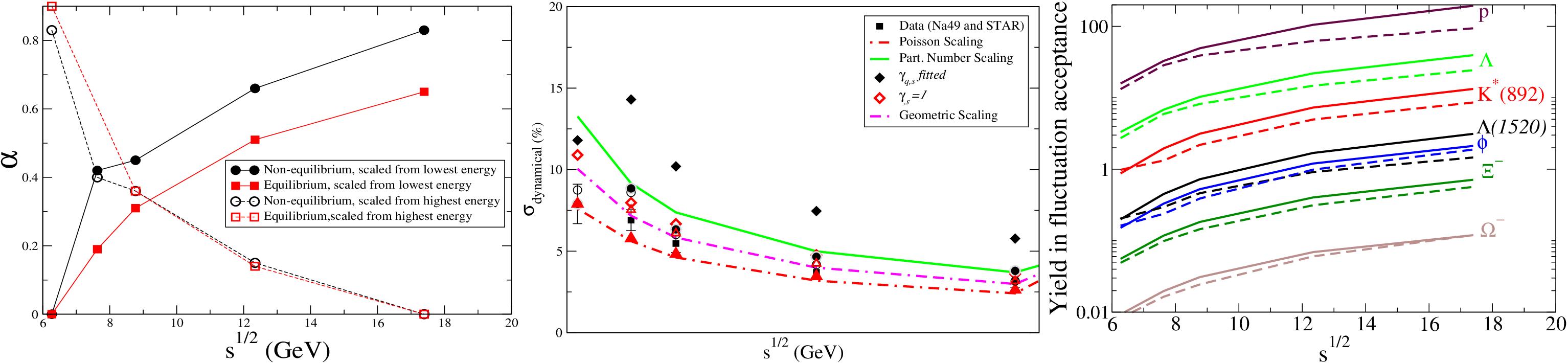}
\caption{\label{kochfigure} Left and center panel: Scaling of $K/\pi$ fluctuations with yields.  The left panel shown the $\alpha$ exponent, defined as in Eq.  \ref{alphaeq}.  Middle panel: The fluctuations observed and calculated with such scaling. Right panel: Yields calculated in the equilibrium and non-equilibrium models assuming the normalization used in the respective calculations corresponds to the physical volume}
\end{center}
\end{figure}

\cite{kochpaper} has done a very useful job of showing that the scaling of fluctuations with $\sqrt{s}$ reflects, to a good approximation, that of yields.  Of course, in the statistical model an {\em exact} scaling of $\sigma_{dyn}^{K/\pi}$ with $\ave{K},\ave{\pi}$ in different energy systems is more complicated than any of the models studied in \cite{kochpaper}, since it reflects changes in $T,\mu$ and possibly $\gamma_{q,s}$ with $\sqrt{s}$.   To try to see how the statistical model fits within the analysis of \cite{kochpaper} we parametrize the best fit set of parameters in \cite{jan_energy,jan_size,jan_search} and \cite{equil_energy} (in the second case $T,\mu_B$,in the first $T,\mu_B,\gamma_{s,q}$ using the exponent $\alpha$
\begin{equation}
\label{alphaeq}
\ave{\pi}^{1/2-\alpha} \ave{K}^{1/2+\alpha} \sigma_{dyn}^{K/\pi} = C =\left. \ave{\pi}^{1/2} \ave{K}^{1/2} \sigma_{dyn}^{K/\pi} \right|_{Reference}
\end{equation}
where $C$ is a constant adjusted from  data (Note that all volume dependence cancels out in the definition of $\alpha$, as expected in the statistical model).  $\alpha$ is simply extracted out of the SHARE estimates \cite{sharev2}
\begin{equation}
\alpha = \frac{ \left.  \ln C - \ln \sigma_{dyn}^{K/\pi} - \left(  \ln \pi + \ln K \right)/2  \right|_{SHARE}}{\left. \ln \ave K- \ln \ave{\pi}\right|_{SHARE}}
\end{equation}
We have performed this procedure with SPS data, taking the parameters from the equilibrium and non-equilibrium values of the analysis in \cite{jan_energy}, and using the $\ave{K}$ and $\ave{\pi}$ expectations from \cite{kochpaper} to fix the normalization.

The result is seen in Fig. \ref{kochfigure} left panel, with full symbols using the lowest SPS energy as reference and empty symbols using the maximum SPS energy.
The conclusions to be drawn from this exercise are two-fold: The first is that none of the scalings examined in \cite{kochpaper} is really compatible with the statistical model.  The rapid change in $\mu_B,\gamma_{q,s}$ really makes the statistical scaling across $\sqrt{s}$ non-trivial.  The second is that the scaling is still close enough to Poissonian that the dependence of the fit parameters on $\sqrt{s}$ washes away if just $K,\pi$ are considered.  
Note that, however, as Fig \ref{kochfigure} shows, these scaled values will be somewhat different from the SHARE prediction.

Note, however, that in this work we did not consider abundances of particles other than $K$,$\pi$.  Since, as shown in \cite{prcfluct}, a yield and a fluctuation are enough to remove any ambiguity between $T$ and $\gamma_q$.
Had the acceptance window considered here been the same for $\ave{K}$ and $\pi$ measurements, this would yield a tight prediction for the value of protons and hyperons.    Taking the normalization constant we fitted with \cite{sharev2} (We want to emphasize this is {\em not} a reliable assumption, we did this for illustrative purposes) we computed the particle yields using the equilibrium and non-equilibrium parameters in \cite{jan_energy,jan_size,jan_search,equil_energy}.  The results are shown in the right hand panel of Fig.\ref{kochfigure}.  As can be seen, a quantitative test of both yields and fluctuations differentiates between the equilibrium and non-equilibrium model much more than a test based on yields and fluctuations alone.    We look forward, therefore, to perform this analysis on data-sets where the acceptance of the two particles in the ratio is the same, and hence protons and hyperons can also be measured within this acceptance.

In conclusion, we have provided an overview of the current results of fluctuations of particle ratios, and tried to interpret them within the framework of the statistical model.   None of the models considered here does a perfect job of describing the data, through models based on Grand-Canonical ensembles better capture the scaling, and the introduction of $\gamma_q$ bring the theory closer to experiment.   A different mixed event subtraction is however needed to confirm these results.
Comparing the effect of resonances inside particle fluctuations with the directly measured resonances shows that the resonance abundance present at chemical freeze-out seems to remain until the final decoupling, in contrast to current estimates from transport models.   While the approximate Poissonian scaling found in the data is expected of the statistical model for all thermal values, the strong variation of $\mu/T$ with $\sqrt{s}$ prevents an exact scaling with energy.  A centrality scan at lower $\sqrt{s}$, as well as a uniform acceptance for both particles in the measured ratio are necessary before the scaling of fluctuations with energy can be unamibiguously interpreted.

This work was supported in part by the Offices of NP 
and HEP within the U.S. DOE Office of Science, the U.S. NSF.
G.T. acknowledges the financial support received from the Helmholtz International
Center for FAIR within the framework of the LOEWE program
(Landesoffensive zur Entwicklung Wissenschaftlich-\"Okonomischer
Exzellenz) launched by the State of Hesse.   
G.W. acknowledges the financial support given by the Alexander Von 
Humboldt foundation research award, and the hospitality provided by the 
JW Goethe Universitat during the time this work was performed.
We thank M.Hauer and Tim Schuster for very helpful discussions, and Tim Schuster and Volker Koch for providing the file used to make Fig. \ref{kochfigure}.


\begin{thebibliography}{15}



\bibitem{Fer50} 
E.~Fermi
{Prog. Theor. Phys.} {\bf 5}, 570 (1950). 
 
 
\bibitem{Pom51} 
I. Pomeranchuk  
{Proc. USSR Academy of Sciences} (in Russian) 
{\bf 43}, 889  (1951).  
 
 
\bibitem{Lan53} 
  LD~Landau, 
  Izv.\ Akad.\ Nauk Ser.\ Fiz.\  {\bf 17}  51-64  (1953). 
 
 
\bibitem{Hag65} 
R. Hagedorn R  
 {Suppl. Nuovo Cimento}  {\bf 2}, 147 (1965). 
 

\bibitem{bdm}
  P.~Braun-Munzinger, D.~Magestro, K.~Redlich and J.~Stachel,
  Phys.\ Lett.\  B {\bf 518}, 41 (2001)
 

\bibitem{equil_energy} 
  J.~Cleymans, H.~Oeschler, K.~Redlich and S.~Wheaton, 
  arXiv:hep-ph/0607164. 
  
\bibitem{jansbook}
~Letessier J,  ~Rafelski J (2002),
Hadrons  quark - gluon plasma,
Cambridge Monogr.\ Part.\ Phys.\ Nucl.\ Phys.\ Cosmol.\  {\bf 18}, 1,
and references therein
 
\bibitem{becattini} 
  F.~Becattini et. al., 
  Phys.\ Rev.\ C {\bf 69}, 024905 (2004) 


\bibitem{nuxu}
  J.~Cleymans, B.~Kampfer, M.~Kaneta, S.~Wheaton and N.~Xu,
  Phys.\ Rev.\  C {\bf 71}, 054901 (2005)

\bibitem{share}
  G.~Torrieri, S.~Steinke, W.~Broniowski, W.~Florkowski, J.~Letessier 
and J.~Rafelski,
  Comput.\ Phys.\ Commun.\  {\bf 167}, 229 (2005)


\bibitem{jaki}
  J.~Noronha-Hostler, C.~Greiner and I.~A.~Shovkovy,
  Phys.\ Rev.\ Lett.\  {\bf 100}, 252301 (2008)

\bibitem{castorina}
  F.~Becattini, P.~Castorina, A.~Milov and H.~Satz,
  arXiv:0911.3026 [hep-ph].


\bibitem{pbmel}
  K.~Redlich, A.~Andronic, F.~Beutler, P.~Braun-Munzinger and J.~Stachel,
  J.\ Phys.\ G {\bf 36}, 064021 (2009)
  [arXiv:0903.1610 [hep-ph]].

\bibitem{prcfluct}
  G.~Torrieri, S.~Jeon and J.~Rafelski,
  Phys.\ Rev.\  C {\bf 74}, 024901 (2006)
  [arXiv:nucl-th/0503026].


\bibitem{sharev2}
  G.~Torrieri, S.~Jeon, J.~Letessier and J.~Rafelski,
  Comput.\ Phys.\ Commun.\  {\bf 175}, 635 (2006)


\bibitem{hagpress}
  R.~Hagedorn,
  Z.\ Phys.\  C {\bf 17}, 265 (1983).



\bibitem{v2fluct}
  S.~Vogel, G.~Torrieri and M.~Bleicher,
  arXiv:nucl-th/0703031.

\bibitem{konch}
  M.~I.~Gorenstein, M.~Hauer, V.~P.~Konchakovski and E.~L.~Bratkovskaya,
  Phys.\ Rev.\  C {\bf 79}, 024907 (2009)
  [arXiv:0811.3089 [nucl-th]].


\bibitem{stephanov}
  M.~A.~Stephanov,
  Phys.\ Rev.\ Lett.\  {\bf 102}, 032301 (2009)
  [arXiv:0809.3450 [hep-ph]].


\bibitem{canonical1}
  S.~Hamieh, K.~Redlich and A.~Tounsi,
  Phys.\ Lett.\  B {\bf 486}, 61 (2000)
  [arXiv:hep-ph/0006024].


\bibitem{canfluct}
  V.~V.~Begun, M.~Gazdzicki, M.~I.~Gorenstein and O.~S.~Zozulya,
  Phys.\ Rev.\  C {\bf 70}, 034901 (2004)

\bibitem{methods}
  C.~Pruneau, S.~Gavin and S.~Voloshin,
  Phys.\ Rev.\  C {\bf 66}, 044904 (2002)
  [arXiv:nucl-ex/0204011].



\bibitem{fluctstar}
        J.~Adams {\it et al.}  [STAR Collaboration],
  Phys.\ Rev.\ Lett.\  {\bf 103}, 92301 (2009)
  [arXiv:0809.3450 [hep-ph]].

\bibitem{na49fluct}
  C.~Alt {\it et al.}  [NA49 Collaboration],
  Phys.\ Rev.\  C {\bf 79}, 044910 (2009)
  [arXiv:0808.1237 [nucl-ex]].


\bibitem{fluctvolume}
  V.~V.~Begun, M.~Gazdzicki and M.~I.~Gorenstein,
  Phys.\ Rev.\  C {\bf 78}, 024904 (2008)
  [arXiv:0804.0075 [hep-ph]].


\bibitem{jeon}
  S.~Jeon and V.~Koch,
  Phys.\ Rev.\ Lett.\  {\bf 83}, 5435 (1999)
  [arXiv:nucl-th/9906074].



\bibitem{fluctreso}
  J.~Adams {\it et al.}  [STAR Collaboration],
  Phys.\ Rev.\ Lett.\  {\bf 97}, 132301 (2006)
  [arXiv:nucl-ex/0604019].


\bibitem{markreso}
  C.~Markert,
  J.\ Phys.\ G {\bf 31}, S169 (2005)
  [arXiv:nucl-ex/0503013].

\bibitem{ratio1}
  G.~Torrieri and J.~Rafelski,
  Phys.\ Lett.\  B {\bf 509}, 239 (2001)
  [arXiv:hep-ph/0103149].

\bibitem{ratio2}
  J.~Rafelski, J.~Letessier and G.~Torrieri,
  Phys.\ Rev.\  C {\bf 64}, 054907 (2001)
  [Erratum-ibid.\  C {\bf 65}, 069902 (2002)]
  [arXiv:nucl-th/0104042].


\bibitem{urqmd1}
  M.~Bleicher and J.~Aichelin,
  Phys.\ Lett.\  B {\bf 530}, 81 (2002)
  [arXiv:hep-ph/0201123].

\bibitem{urqmd2}
  S.~Vogel, J.~Aichelin and M.~Bleicher,
  arXiv:0908.3811 [hep-ph].

\bibitem{urqmd3}
  S.~Haussler, Fluctuations in ultra-relativistic heavy ion collisions from Microscopic descriptions, PhD thesis


\bibitem{qm2007}
  G.~Torrieri,
  Int.\ J.\ Mod.\ Phys.\  E {\bf 16}, 1783 (2007)
  [arXiv:nucl-th/0702062].

\bibitem{starscan}
  G.~Odyniec  [STAR Collaboration],
  J.\ Phys.\ G {\bf 35}, 104164 (2008).

\bibitem{phescan}
  T.~Sakaguchi and f.~t.~P.~collaboration,
  arXiv:0908.3655 [hep-ex].

\bibitem{nica}
  A.~N.~Sissakian, A.~S.~Sorin and V.~D.~Toneev,
  Phys.\ Part.\ Nucl.\  {\bf 39} (2008) 1062.


\bibitem{kochpaper}
  V.~Koch and T.~Schuster,
  arXiv:0911.1160 [nucl-th].


\bibitem{jan_energy}
  J.~Letessier and J.~Rafelski,
  Eur.\ Phys.\ J.\  A {\bf 35}, 221 (2008)
  [arXiv:nucl-th/0504028].

\bibitem{jan_size}
  J.~Rafelski, J.~Letessier and G.~Torrieri,
  Phys.\ Rev.\  C {\bf 72}, 024905 (2005)
  [arXiv:nucl-th/0412072].

\bibitem{jan_search}
  G.~Torrieri and J.~Rafelski,
  New J.\ Phys.\  {\bf 3}, 12 (2001)
  [arXiv:hep-ph/0012102].






\end{thebibliography}
\end{document}